\documentclass[amsmath,amssymb,reprint,superscriptaddress]{revtex4-2}

\usepackage{graphicx}
\usepackage{color}
\usepackage{url}
\usepackage{accents}
\usepackage{braket}
\usepackage{algpseudocode}
\usepackage{algorithm}

\def\mathi{{\mathrm i}}

\newcommand{\Nb}{\ensuremath{{N_\mathrm{bath}}}}
\newcommand{\Nbinit}{\ensuremath{{N_\mathrm{bath}^0}}}
\newcommand{\Nimp}{\ensuremath{{N_\mathrm{SO}}}}
\newcommand{\wmax}{\ensuremath{{\omega_\mathrm{max}}}}
\newcommand{\Nbgen}{\ensuremath{{\tilde{N}_\mathrm{b}}}}

\newcommand{\Vgen}{\ensuremath{{\tilde{V}}}}

\newcommand{\vecv}{\ensuremath{{\boldsymbol{v}}}}

\newcommand{\deltath}{\ensuremath{{\delta_\mathrm{th}}}}

\newcommand{\bx}{\ensuremath{{\boldsymbol{x}}}}
\newcommand{\by}{\ensuremath{{\boldsymbol{y}}}}
\newcommand{\bA}{\ensuremath{{\boldsymbol{A}}}}

\begin{document}

\title{
Sparse modeling of large-scale quantum impurity models with low symmetries
}

\author{Hiroshi Shinaoka}
\affiliation{Department of Physics, Saitama University, Saitama 338-8570, Japan}
\email{shinaoka@mail.saitama-u.ac.jp}

\author{Yuki Nagai}
\affiliation{CCSE, Japan Atomic Energy Agency, 178-4-4, Wakashiba, Kashiwa, Chiba, 277-0871, Japan}
\affiliation{
Mathematical Science Team, RIKEN Center for Advanced Intelligence Project (AIP), 1-4-1 Nihonbashi, Chuo-ku, Tokyo 103-0027, Japan
}

\begin{abstract}
Quantum embedding theories provide a feasible route for obtaining quantitative descriptions of correlated materials.
However, a critical challenge is solving an effective impurity model of correlated orbitals embedded in an electron bath.
Many advanced impurity solvers require the approximation of a bath continuum using a finite number of bath levels, producing a highly nonconvex, ill-conditioned inverse problem.
To address this drawback, this study proposes an efficient fitting algorithm for matrix-valued hybridization functions based on a data-science approach, sparse modeling, and a compact representation of Matsubara Green's functions.
The efficiency of the proposed method is demonstrated by fitting random hybridization functions with large off-diagonal elements as well as those of a 20-orbital impurity model for a high-$T_\mathrm{c}$ compound, LaAsFeO, at low temperatures ($T$).
The results set quantitative goals for the future development of impurity solvers toward quantum embedding simulations of complex correlated materials.
\end{abstract}

\maketitle
Simulating correlated materials is one of the major challenges in the field of condensed matter physics. Local density approximation (LDA) based on the density functional theory has achieved significant success in describing the ground-state properties of many weakly correlated materials. However, LDA fails to describe correlated materials such as Mott insulators and high-$T_{c}$ superconductors.
A naive direct simulation of the first-principles Hamiltonians of correlated materials is not feasible, owing to the exponential scaling of its required computational resources. 

In recent times, extensive efforts have been made to use Greens-function-based quantum embedding theories to simulate correlated materials.
Quantum embedding theories circumvent the need for exponential scaling by mapping the entire computationally intractable system onto an auxiliary impurity model of correlated orbitals embedded in a bath of noninteracting electrons. Examples of this methodology are the dynamical mean-field theory (DMFT)~\cite{Georges:1996un}, $GW$+DMFT~\cite{Biermann02}, nonlocal extensions of DMFT~\cite{Galler:2017bh,Rohringer2018}, and the self-energy embedding theory~\cite{Kananenka:2015,Iskakov2020}.

The limitation of these methodologies is in solving impurity models. Quantitative descriptions of correlated materials, such as predicting $T_\mathrm{c}$,
will require the solution of an impurity model with several spin orbitals ($\Nimp \gg 10$) and low symmetries. 
Over the past few years, various sophisticated impurity solvers have been developed.
Some examples include impurity solvers based on truncated exact diagonalization~\cite{Lu:2014de,Go:2017fe}, configuration interaction~\cite{Zgid:2012ck,MejutoZaera2019}, coupled-cluster theory~\cite{Shee2019,Zhu2019}, matrix product states~\cite{Alexander2015,Linden:2020iu} and tensor networks~\cite{Bauernfeind:2017du}. 
Now, these state-of-the-art algorithms even allow the handling of a few correlated atoms~\cite{Alexander2015}.

\begin{figure}[t]
  \centering
  \includegraphics[width=0.9\linewidth]{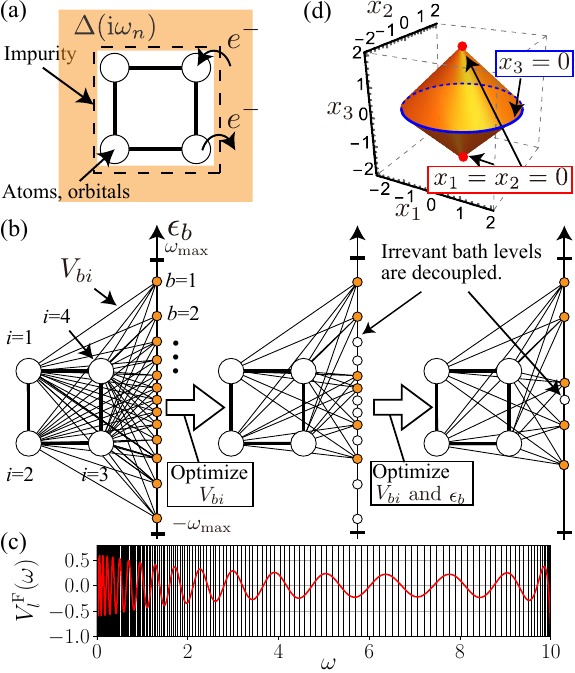}
  \caption{
  (Color online) (a) Quantum impurity model. The bath is represented by a hybridization function $\Delta(\mathi \omega_n)$.
  (b) Two-step optimization procedure of the present algorithm.
  (c) Highest-order IR basis functions $V_{N_l-1}^\mathrm{F}(\omega)$ (red solid curve) and the positions of the initial $\epsilon_b$ (bold and thin solid vertical lines) for $\beta=100$, $\wmax=10$, $N_l=72$ and $N_\mathrm{div}=5$ [see (b) and the text].
  (d) Isosurface of the group LASSO regularization term in Eq.~(\ref{eq:groupLASSO}).
  }
  \label{fig:method}
\end{figure}

All the above-mentioned impurity solvers rely on approximating a bath continuum with a finite number ($\Nb$) of bath levels.
Self-consistent calculations to determine a bath are stably performed in Matsubara frequencies.
However, fitting a hybridization function, which represents a bath, is an ill-conditioned and nonconvex inverse problem~\cite{Koch:2008bx,Senechal2010,semidefinite2020}.
This becomes more severe for multiple spin orbitals and low symmetries. Thus, an approach to discretize a hybridization function for $\Nimp\gg 10$ and the size of $\Nb$ required for an accurate approximation are yet to be clarified.
As a result, advanced impurity solvers cannot be fully utilized.

It is noted that another promising route to solving much larger-scale quantum impurity models is \textit{quantum} computing~\cite{Bauer:2016fc,Rubin2016,Ma2020}.
Very recently, preliminary calculations for a single-orbital impurity model have been performed using an IBM quantum computer without fault tolerance~\cite{Runggerpreprint,keen2019quantumclassical}.
As the number of available qubits increases and noise levels decrease, quantum algorithms may begin to compete with or supersede classical algorithms.
However, the currently proposed algorithms for quantum computers rely on discretizing the bath continuum.
Thus, the bath discretization will remain to be a critical step of quantum embedding simulations.

It was recently proven that $\Nb$ ($\propto \Nimp$) suffices to compute exact ground-state energies based on removing redundant bath degrees of freedom using basis rotation of spin orbitals and bath levels~\cite{Bravyi:2017cc}.
This basis rotation however generates nonlocal Coulomb interactions, which is not preferable for some impurity solvers.
Furthermore, the argument is limited to zero $T$.
An interesting question is if one can construct such a compact dicretized model by fitting the hybridization function without mixing the local and bath degrees of freedom for finite $T$.

This letter proposes the use of a data-science-based approach, sparse modeling~\cite{Elad,Otsuki:2017er}, to achieve efficient discretization of large-scale impurity models with low symmetries.
Its novelty lies in (i) compactification of discretized models through automatic selection of relevant bath levels by sparse modeling, and (ii) projection to a recently proposed compact basis in Matsubara frequencies, the intermediate representation (IR) basis~\cite{Shinaoka:2017ix,Chikano:2018gd,irbasis2019,Otsuki:2020fn}.
The efficiency of the proposed algorithm is demonstrated with random hybridization functions for $\Nimp\gg 10$ with low symmetries.
It is shown that for a fixed fitting tolerance,
$\Nb\propto \Nimp\ln (\beta W)$, where $W$ is the spectral width and $\beta$ is inverse temperature.
Additionally, a realistic five-orbital 2$\times$ 2 cluster impurity model for a Fe-based high-$T_\mathrm{c}$ superconductor LaFeAsO is analyzed, and $\Nb$ is estimated.

\textit{General impurity model}--
A general impurity model is defined by the action
\begin{align}
   S_\mathrm{imp} &= \int_0^\beta d\tau \mathcal{H}_\mathrm{loc}(\tau) \nonumber\\
    & + \int_0^\beta d\tau d\tau^\prime \sum_{i,j=1}^\Nimp c^\dagger_i (\tau) \Delta_{ij}(\tau-\tau^\prime) c_j(\tau^\prime),
\end{align}
where $c^\dagger_i$ and $c_i$ are Grassmann variables for the $i$-th spin orbital in the impurity.
The local action $\mathcal{H}_\mathrm{loc}$ acts on the impurity, while all the bath/environment information is encoded in the matrix-valued hybridization function $\Delta_{ij}(\tau)$.
If $\mathcal{H}_\mathrm{loc}$ is instantaneous in $\tau$, 
this action can be regarded as the result of integrating out auxiliary degrees of freedom from a Hamiltonian model,
\begin{align}
   \mathcal{H}_\mathrm{imp} &= \mathcal{H}_\mathrm{loc} + \sum_{b=1}^\Nb \epsilon_b \hat{d}^\dagger_b \hat{d}_b + \sum_{i=1}^\Nimp\sum_{b=1}^\Nb (V_{bi} \hat{d}_b^\dagger \hat{c}_i + V_{bi}^* \hat{c}^\dagger_i \hat{d}_b),
\end{align}
where $\hat{c}$ and $\hat{d}$ are annihilation operators acting on spin orbitals and bath levels, respectively, and $V_{bi}$ and $\epsilon_b$ are bath parameters.
The bath levels are assumed to be sufficient to precisely satisfy the equality condition $\Delta_{ij}(\mathi \omega_n) = \sum_b \frac{V_{bi}^*V_{bj}}{\mathi \omega_n - \epsilon_b}$.
Each bath level is hybridized with all the spin orbitals in the impurity [see Fig.~\ref{fig:method}(b)].
Note that this transformation is not unique.

\textit{Number of parameters required to represent a bath}--
We expand a hybridization function as
\begin{align}
  \Delta_{ij}(\mathi \omega_n) &= \sum_{l=0}^\infty \Delta_{ij}(l) U_l^\mathrm{F}(\mathi \omega_n)~~~\mathrm{with}~|\Delta_{ij}(l)| \propto S_l^\mathrm{F},
\end{align}
where fermionic IR basis functions $U_l^\mathrm{F}(\mathi\omega_n)$ depend on $\beta$ and a cutoff frequency $\wmax$ for spectral functions~\cite{Shinaoka:2017ix} (the notation used in Ref.~\onlinecite{irbasis2019} is utilized herein).
$S_l^\mathrm{F}$ denote the singular values of the kernel in the Lehmann representation, being system independent.
Because $S_l^\mathrm{F}$ decay super-exponentially,
the summation is truncated at $S_{N_l-1}^\mathrm{F}/S_0^\mathrm{F} \simeq 10^{-15}$.
$N_l$ grows only logarithmically with $\beta\wmax$ ($N_l$=40, 72, 104 for $\beta\wmax=10^2,10^3, 10^4$, respectively).
This implies that any bath can be represented using $\Nimp^2 N_l~(\propto \Nimp^2 \ln \beta)$ parameters alone, regardless of the number of physical degrees of freedom encoded in it.
On the other hand, a discretized model involves $\Nb$ + $\Nimp\Nb$ parameters.
This implies that $\Nb$ must scale at least as $\mathcal{O}(\Nimp \ln \beta)$.

\textit{Efficient discretization algorithm}--
The hybridization fitting is a highly non-convex and ill-conditioned inverse problem.
To alleviate this,
we consider the regularized cost function
\begin{align}
    &f(\boldsymbol{x})  = \sum_{n=-\infty}^{\infty}\sum_{i,j=1}^\Nimp\left|\Delta_{ij}(\mathi \omega_n) - \sum_{b=1}^\Nbinit \frac{V_{ib}^* V_{bj}}{\mathi \omega_n - \epsilon_b}\right|^2
   + \alpha \sum_{b=1}^\Nbinit \left\| \vecv_b\right\|\label{eq:cost-func-wn}\\
    & = \sum_{l=0}^{N_l-1}\sum_{i,j=1}^\Nimp\left|\Delta_{ij}(l) + \sum_{b=1}^\Nbinit V_{ib}^* V_{bj} S_l^\mathrm{F} V_l^\mathrm{F}(\epsilon_b)\right|^2 + \alpha \sum_{b=1}^\Nbinit \left\| \vecv_b \right\|
    \label{eq:cost-func-IR}
\end{align}
where $\Nbinit$ is the number of initial bath levels, and $\|\cdots\|$ denotes the Frobenius norm,
$\vecv_b \equiv (V_{b1},\cdots,V_{b\Nimp})^\mathrm{T}$, $\alpha>0$.
The fitting parameters are $\boldsymbol{\epsilon} = (\epsilon_1, \cdots )$ and $\boldsymbol{V}=(V_{11}, \cdots, V_{1\Nimp},\cdots)$.
The second term serves to prune redundant bath levels, as will be explained further on.
In Eq.~(\ref{eq:cost-func-IR}), the truncation error in Matsubara frequencies is eliminated by transforming the cost function to the IR basis.
$V^\mathrm{F}_l(\omega)$ are IR basis functions defined in $[-\wmax, \wmax]$~\cite{Shinaoka:2017ix}.
Hereinafter, only cases where $V_{bi}$ are real are considered.

Minimizing Eq.~(\ref{eq:cost-func-IR}) with $\alpha=0$ would yield an unfavorable solution in which all bath levels would be strongly or weakly coupled to the impurity as $\left\| \vecv_b \right\| \neq 0$. Thus, we prefer a \textit{sparse solution}, in which $\left\| \vecv_b \right\|=0$ for irrelevant bath levels. This is achieved by regularization, which is based on the so-called group least absolute shrinkage and selection operator (group LASSO)~\cite{Yuan2006}.

To understand how the group LASSO works,
a general underdetermined linear regression problem $\boldsymbol{y}=\boldsymbol{A}\boldsymbol{x}$ is considered,
in which $\boldsymbol{x}$ (data to be fitted) and $\boldsymbol{y}$ (fitting parameters) are vectors of $N$- and $M$-dimensions, respectively.
$\boldsymbol{A}$ is an $M\times N$ coefficient matrix ($M<N$).
This system has an infinite number of solutions.
Its degeneracy is lifted using a group LASSO regularization term,
\begin{align}
	\bx^* = \underset{\bx}{\mathrm{argmin}}~\|\boldsymbol{y}-\boldsymbol{A}\boldsymbol{x}\|^2 + \alpha \left(\sqrt{x_1^2 + x_2^2} + \sqrt{x_3^2}\right),\label{eq:groupLASSO}
\end{align}
where $N=3$ and $\boldsymbol{x}=(x_1, x_2, x_3)^\mathrm{T}$ are taken for simplicity.
Note that $x_1$ and $x_2$ are grouped, while $x_3$ forms another group on its own.
This regularization term is the sum of Frobenius norms of vectors consisting of fitting parameters in each group.
As shown in Fig.~\ref{fig:method}(d), an isosurface of this term has sharp corners and edges, where grouped fitting parameters are either entirely zero or entirely nonzero.
The solution of Eq.~(\ref{eq:groupLASSO}) represents the contact point(s) between an isosurface and the plane $\by=\bA \bx$.
Because these solutions lie somewhere on the sharp corners and edges, the group LASSO removes irrelevant fitting parameters in a grouped manner.
In the present study, the group LASSO decouples an irrelevant bath level $b$ by individually removing all $\Nimp$ coupling constants $V_{ib}$.

Following  the above argument on the scaling of $\Nb$, 
we prepare $\mathcal{O}(\Nimp \ln \beta)$ poles $\epsilon_b$
based on the distribution of the roots of the highest-order basis function $V^\mathrm{F}_{N_l-1}(\omega)$.
$V^\mathrm{F}_l(\omega)$ has $l$ roots in the interval of $[-\wmax, \wmax]$, and they are nonuniformly distributed among real frequencies [see Fig.~\ref{fig:method}(c)].
The cost function (\ref{eq:cost-func-IR}) depends on $\epsilon_b$ through the IR basis functions alone.
If $\epsilon_b$ lies between the wide interval between two neighboring roots, particularly for high frequencies,
the value of the cost function is insensitive to 
a shift of $\epsilon_b$ in this interval,
which is a source of the ill-condition.
To alleviate this problem, $N_\mathrm{bath}^0$ and $\epsilon_b$ are chosen as follows.
First, a grid consisting of the boundary points $\pm \wmax$ and the roots is constructed.
The middle points of two neighboring grid points define the coarser grid shown in Fig.~\ref{fig:method}(c).
$\epsilon_b$ are initialized to these coarse grid points and  $N_\mathrm{div}$ equal division points of each interval of the coarse grid~\footnote{This method of constructing a sparse grid is inspired by the sparse sampling method proposed in Ref.~\onlinecite{Li:2020eu}}.
$N_\mathrm{div}=c\Nimp$ ($c$=2--10) is taken such that 
$\Nbinit \gg \Nb$ after optimization.

The bath parameters $\boldsymbol{V}$ and $\boldsymbol{\epsilon}$ are optimized using the two-step procedure [Fig.~\ref{fig:method}(b)].
First, only $\boldsymbol{V}$ is optimized with $\boldsymbol{\epsilon}$ fixed at the initial estimate.
Although this is still a non-convex optimization, we found that this is empirically stable for $\alpha>0$.
The regularization term, based on the so-called group LASSO~\cite{Yuan2006}, suppresses $\|\vecv_b\|$ of some of the bath levels to zero, while maintaining the finiteness of $\|\vecv_b\|$ of the rest.
After convergence, the bath levels that are almost decoupled from the impurity are eliminated.
In the second step, $\boldsymbol{V}$ and $\boldsymbol{\epsilon}$ are simultaneously optimized, which generally reduces the value of the cost function \textit{less} significantly than the first step.
Finally, redundant bath levels are removed again.
A quasi-Newton method is used for optimization~\footnote{ 
It was found that, in practice, the nonanalytic nature of the cost function at $\vecv_b=0$ is not significant.}.
The algorithm is detailed in Supplemental Material.

\begin{figure}[h!]
  \centering
  \includegraphics[width=0.9\linewidth]{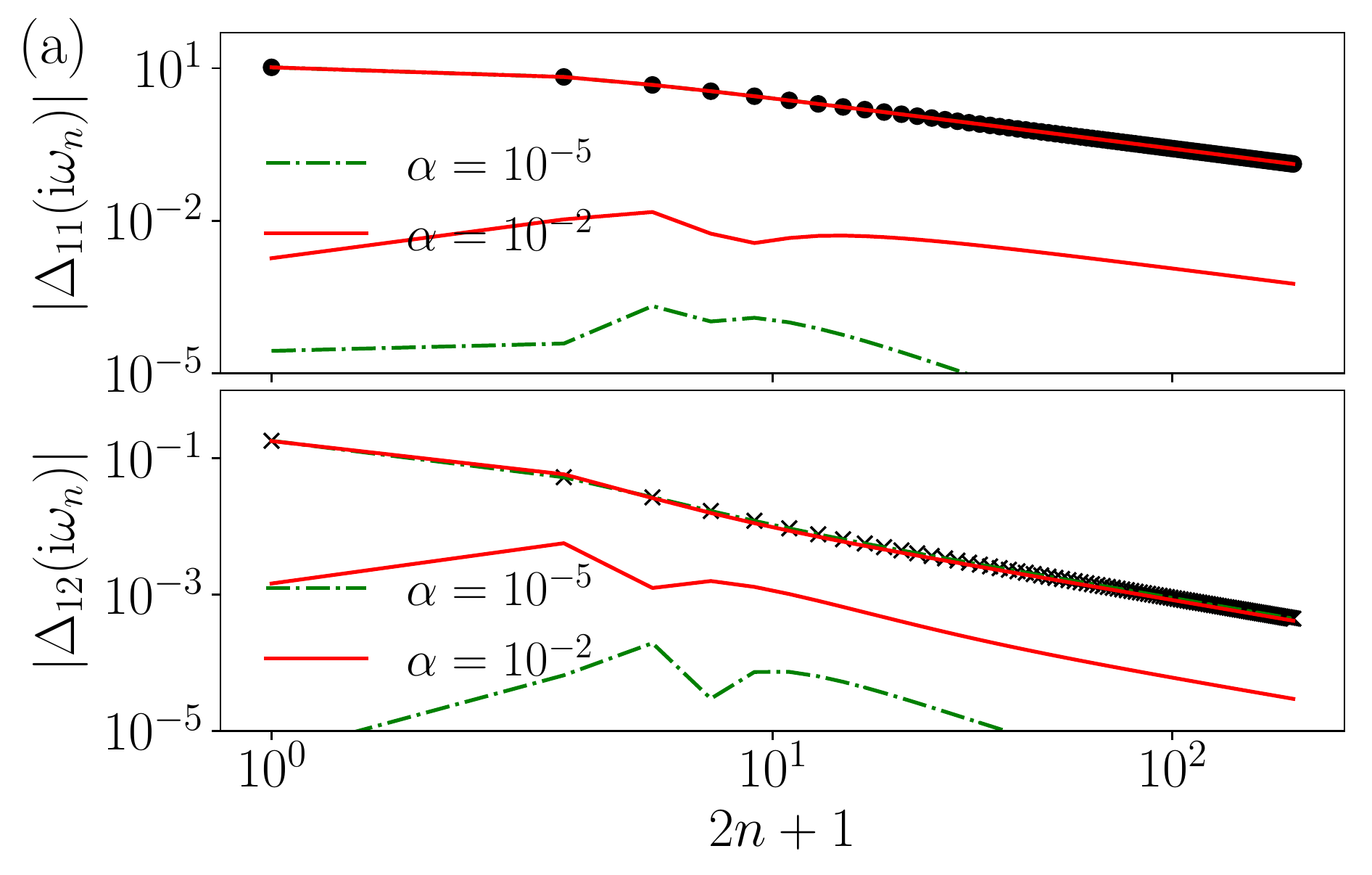}
    \includegraphics[width=0.9\linewidth]{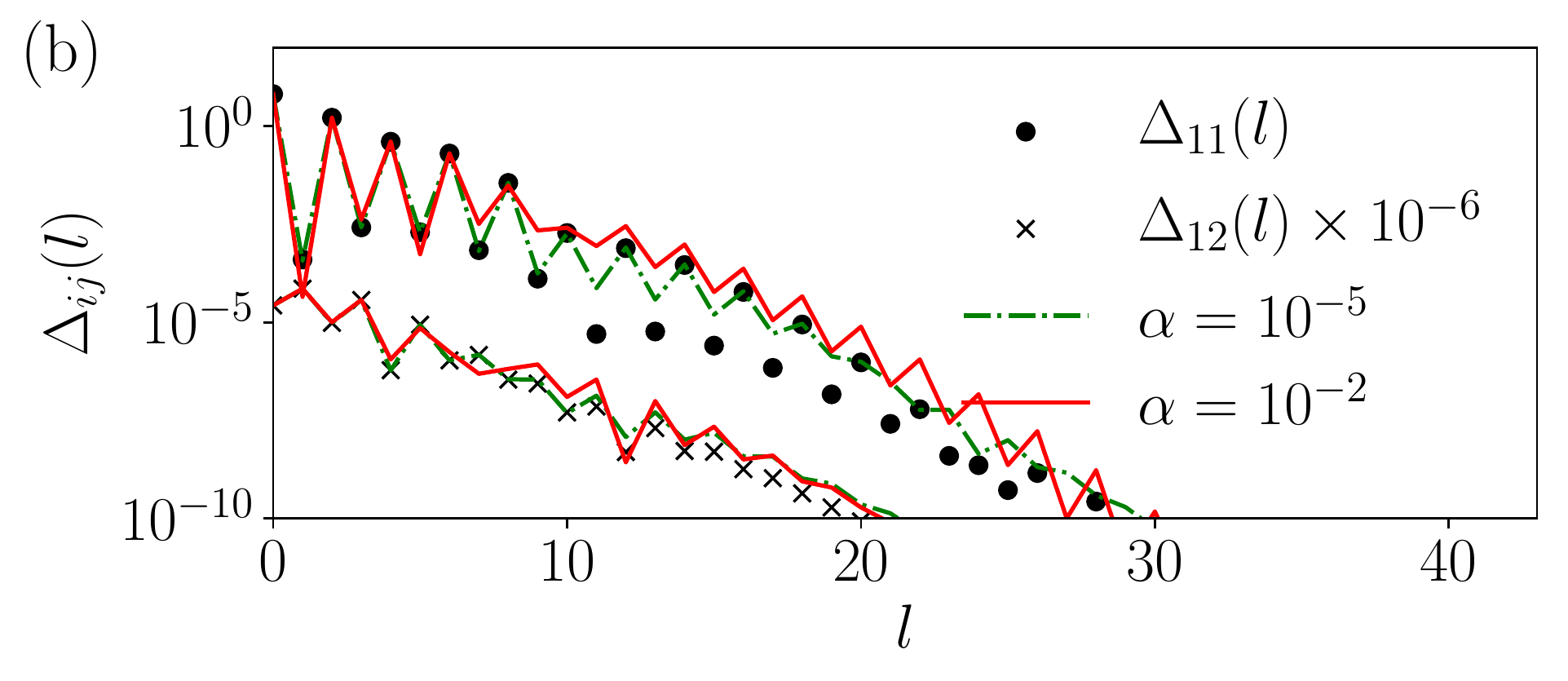}
  \includegraphics[width=0.9\linewidth]{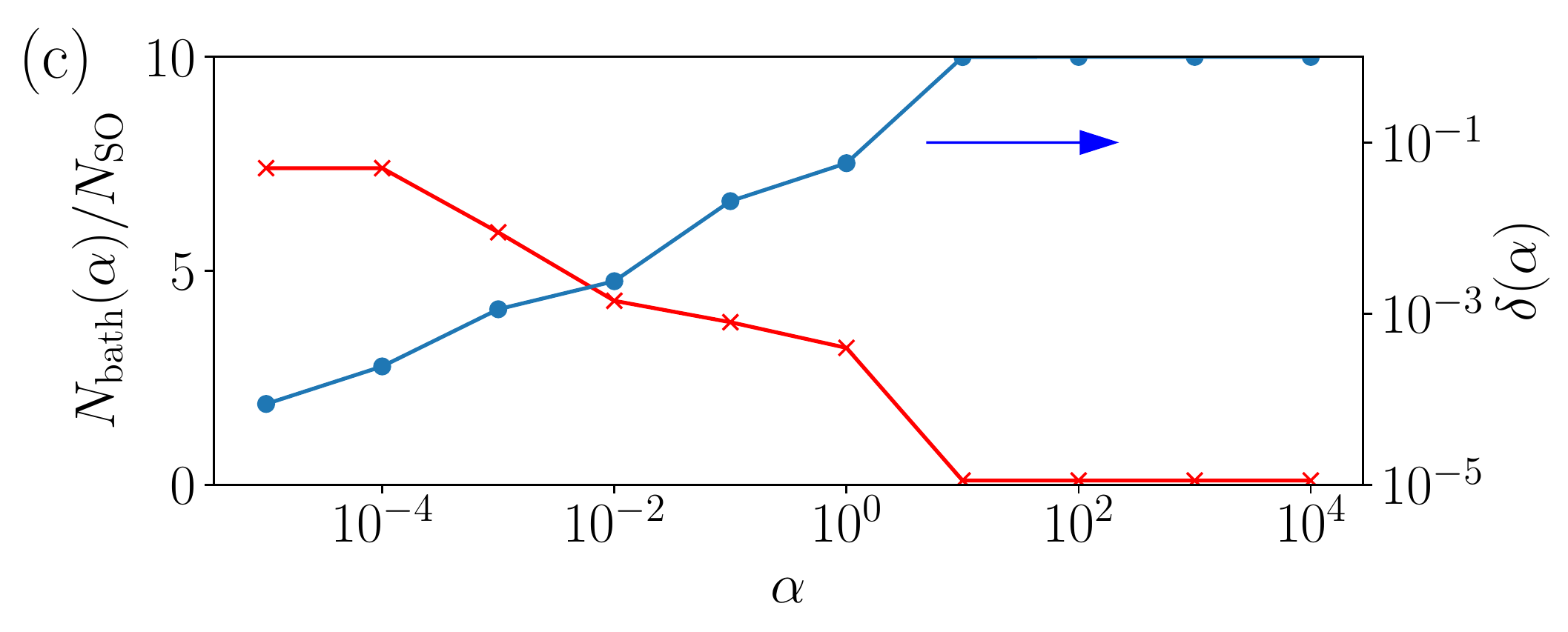}
  \includegraphics[width=0.9\linewidth]{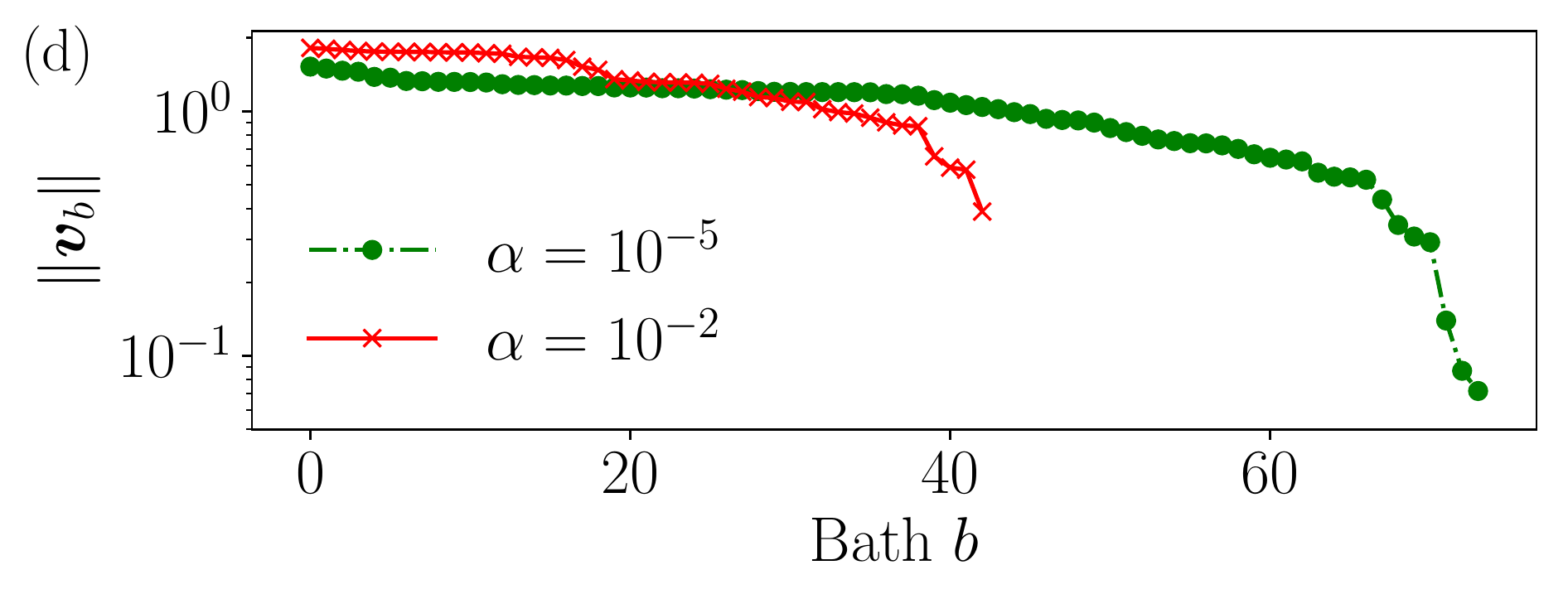}
  \includegraphics[width=0.9\linewidth]{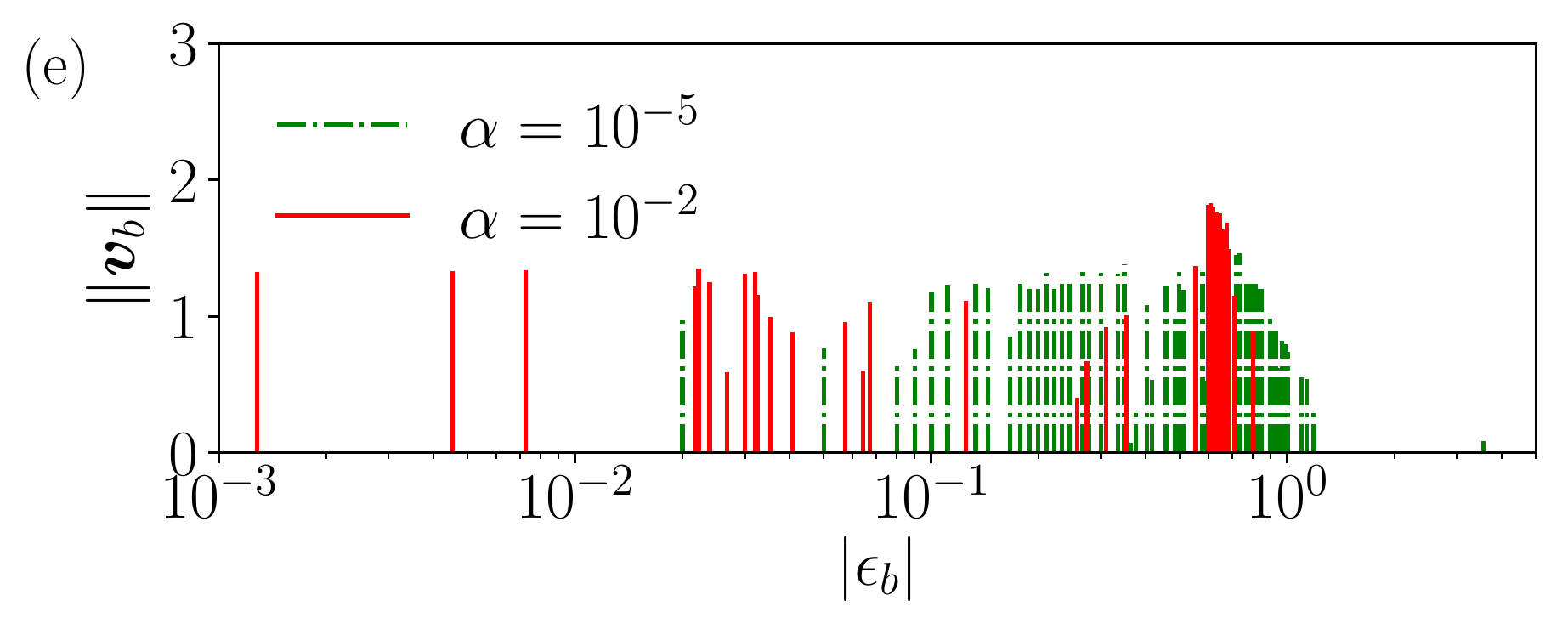}
  \vspace{-1em}
  \caption{
  (Color online) 
  Results of a sample for $\Nimp=10$ and $\beta=10$.
  The hybridization function and the fitted results are compared in Matsubara frequencies [(a)] and in the IR basis [(b)], respectively.
  In (a) and (b), the fitting error is also shown.
  (c) Number of bath levels $\Nb$ and relative residual norm $\delta$.
  (d) Strength of the coupling to the impurity $\|\vecv_b\|$ in decreasing order. (e) Positions of $\epsilon_b$ and $\|\vecv_b\|$.
  }
  \label{fig:random}
\end{figure}
\begin{figure}
  \centering
  \includegraphics[width=1.0\linewidth]{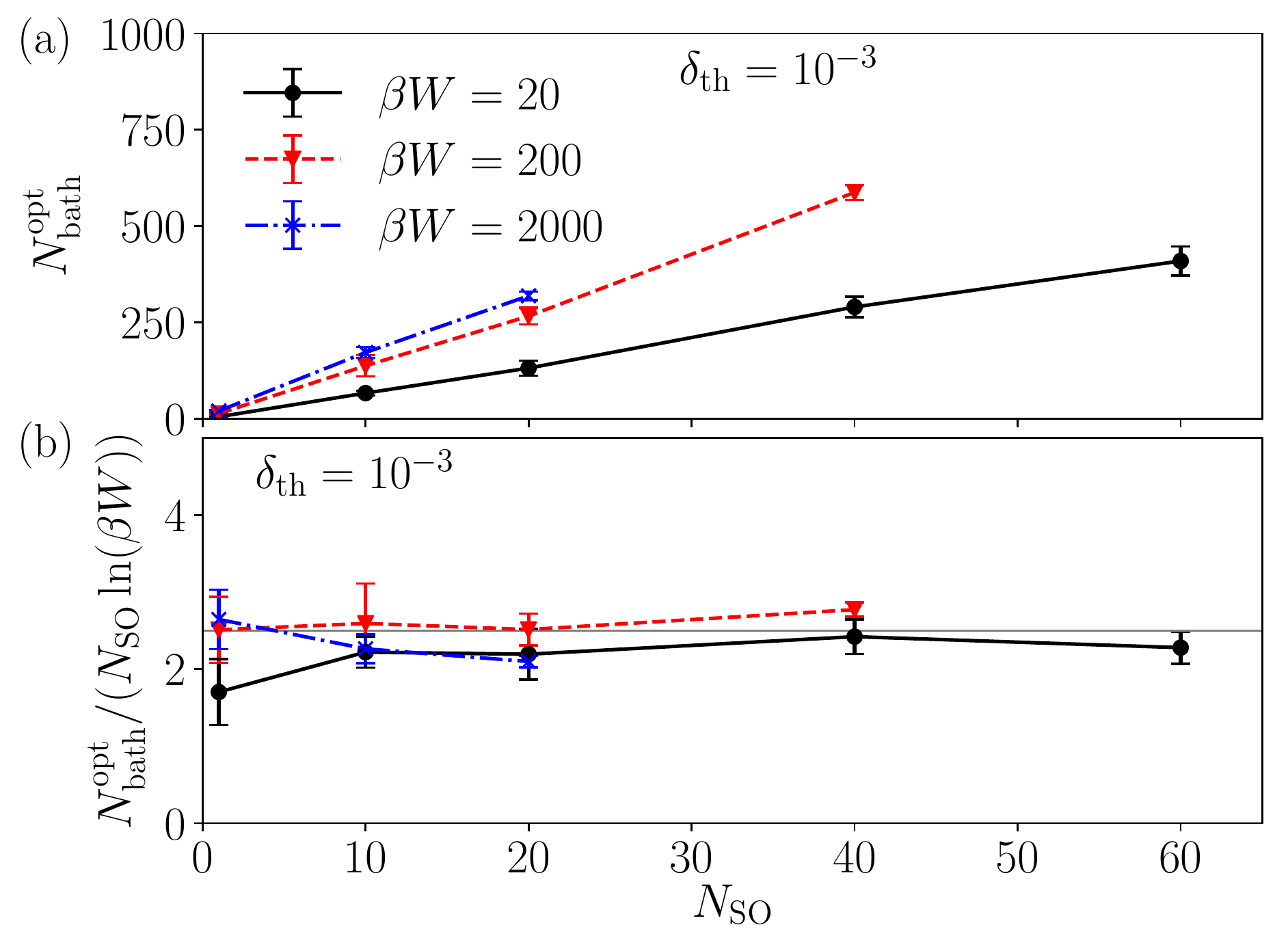}
  \includegraphics[width=1.0\linewidth]{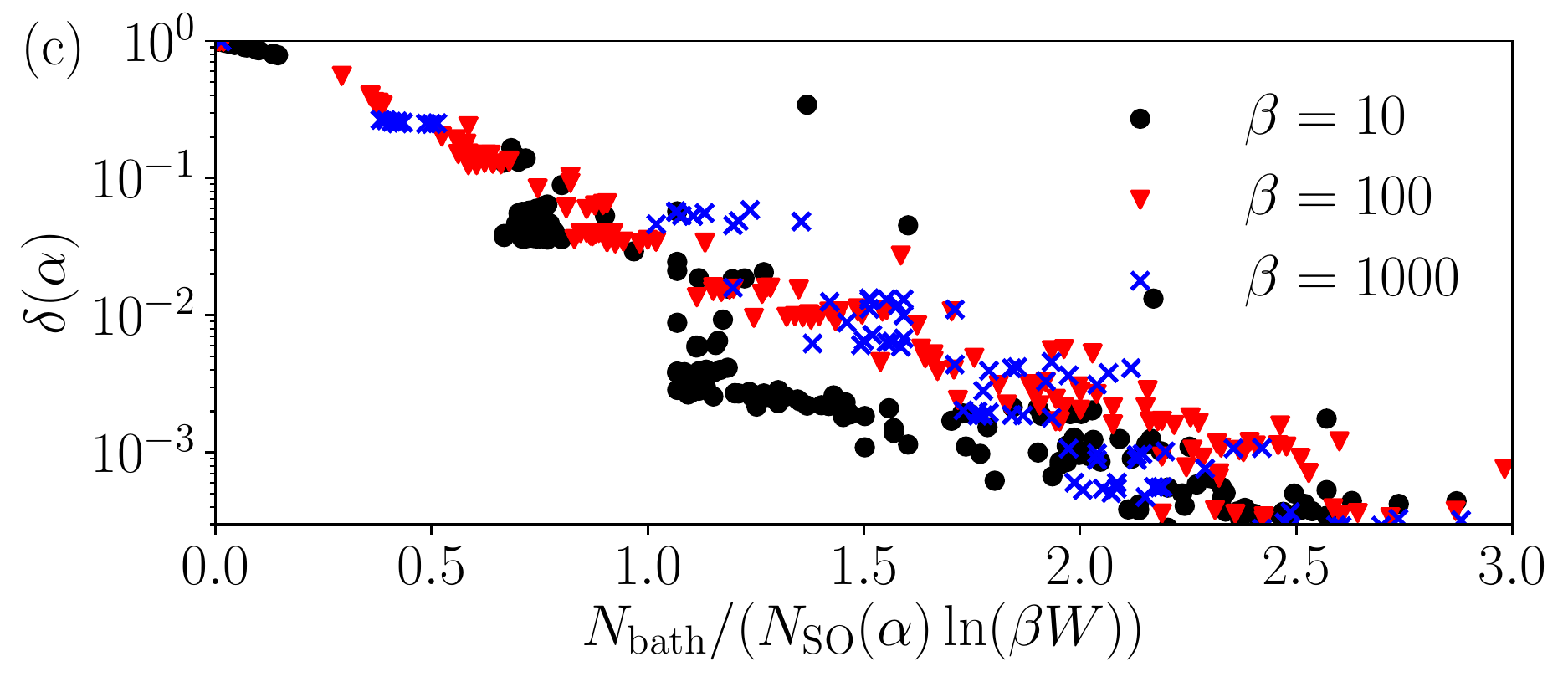}
  \caption{
  (Color online) 
  (a) Number of bath levels $\Nb$ required to fit the random models within the threshold of $\deltath=10^{-3}$.
  The error bars denote the standard deviations computed with ten samples.
  (b) Scaling plot of the same data as in (a).
  The horizontal line serves as a visual guide.
  (c) Scaling plot of data obtained for all values of $\Nimp$, $\beta$, and $\alpha$.
  The fitting error $\delta$ vanishes exponentially with increasing $\Nb$.
  }
  \label{fig:scaling}
\end{figure}

\textit{Results of random models with low symmetries}--
First, the present algorithm was benchmarked for 
an ensemble of random matrix-valued hybridization functions generated as
    $\Delta_{ij}(\mathi \omega_n) = \frac{1}{\sqrt{\Nbgen}}\sum_{b=1}^{\Nbgen}\frac{\Vgen_{bi}^* \Vgen_{bj}}{\mathi \omega_n - \tilde{\epsilon}_b},$
where $\Vgen_{bj}$ are drawn from a uniform distribution on $[-1/2,1/2]$,
while $\tilde{\epsilon}_b$ are uniformly and densely distributed in the interval of $[-W/2,W/2]$ with the full spectral width $W=2$.
$\Nbgen$ was taken to be sufficiently large, $\Nbgen=10N^2_\mathrm{SO}\beta$.
This model represents gapless and quasi-continuous baths with low symmetries.
We took $\wmax=10$ ($\gg W/2$) and introduced initial bath levels in $[-\wmax, \wmax]$ to test the algorithm’s ability to remove redundant bath levels [see Fig.~\ref{fig:method}(c)].
The hybridization functions were transformed into the IR basis before fitting.

Figure~\ref{fig:random}(a) and ~\ref{fig:random}(b) show a sample for $\Nimp=10$ and $\beta=10$ in the Matsubara frequencies and the IR basis, respectively.
The hybridization function has large off-diagonal components. 
In the IR basis, the coefficients decay exponentially with $l$, as in the singular values $S^\mathrm{F}_l$.
The relevant information in the hybridization function is compactly represented by a few IR coefficients.

These hybridization functions were then fitted.
Figure~\ref{fig:random}(c) shows the results obtained for a wide range of $\alpha$.
For sufficiently large $\alpha$ values, a trivial solution is obtained with $\Nb\simeq 0$, where almost all bath levels are removed. 
As $\alpha$ decreases, the relative residual norm also decreases, and more bath levels remain in the solution.
A minimal model for a desired fitting accuracy can be obtained by simply varying $\alpha$.

Figure~\ref{fig:random}(d) shows $\|\vecv_b\|$ obtained for $\alpha=10^{-2}$ and $10^{-5}$.
In both cases, $\|\vecv_b\|$ plateaus and then drops steeply owing to the regularization.
As shown in Fig.~\ref{fig:random}(e),
the distributions of $\epsilon_b$ for the two values of $\alpha$ are nonuniform.
The regularization removes the majority of the redundant bath levels in $W/2~(=1) < |\epsilon_b| < \wmax~(=10)$.

Following this, the quality of the fits were assessed.
As seen in Fig.~\ref{fig:random}(b),
the fitted model reproduces the complex structures of $\Delta_{ij}(l)$ up to $l\simeq 8$.
Figure~\ref{fig:random} (a) shows that in Matsubara frequencies, the hybridization function is well-fitted from low to high frequencies, and it does not exhibit overfitting.

Then, the $\Nimp$ and $\beta$ dependencies were investigated.
By changing $\alpha$, we can estimate the minimum number of bath levels
$N_\mathrm{bath}^\mathrm{opt}$ required to reach the threshold $\deltath=10^{-3}$ for the relative residual norm.
Ten samples were taken at each parameter.
Figure~\ref{fig:scaling}(a) shows that 
the required $\Nb$ grows approximately linearly with $\Nimp$ for a fixed $\beta$,
whereas it grows slowly with $\beta$.
The scaling plot in Fig.~\ref{fig:scaling}(b) strongly supports the expected scaling relation
\begin{align}
   N_\mathrm{bath}^\mathrm{opt}(\deltath) &\simeq c \Nimp \ln \beta W\label{eq:scaling}
\end{align}
with $c\simeq 2.5$.
Figure~\ref{fig:scaling}(c) plots the data points for all values of $\alpha$, $\beta$, and $\Nimp$.
The error vanishes exponentially with $\Nb$, indicating
the logarithmic dependence of $c$ on $\deltath$ in Eq.~(\ref{eq:scaling}).
For quantum embedding simulations at zero $T$, $\beta$ in Eq.~(\ref{eq:scaling}) is replaced by $1/T_\mathrm{fict}$ where $T_\mathrm{fict}$ is a small fictitious temperature that sets the energy resolution of the simulations.
%

\begin{figure}[h!]
  \centering
  \includegraphics[width=\linewidth]{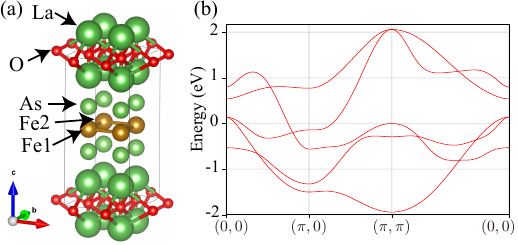}
  \includegraphics[width=\linewidth]{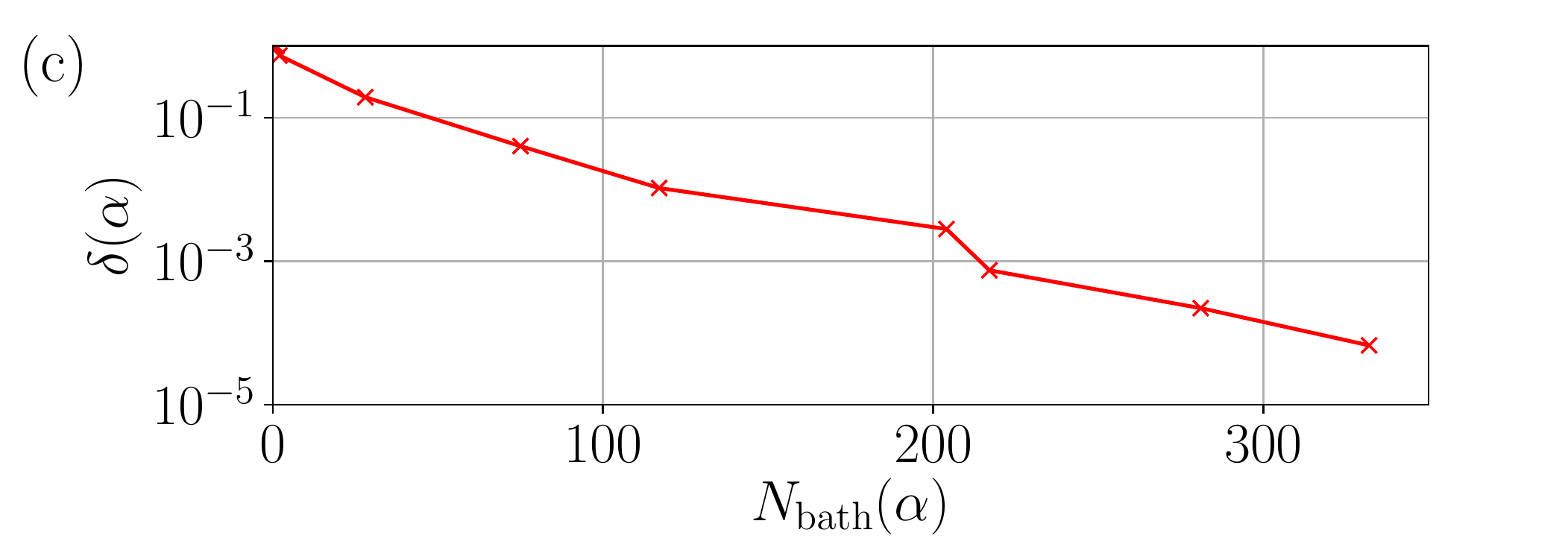}
  \includegraphics[width=\linewidth]{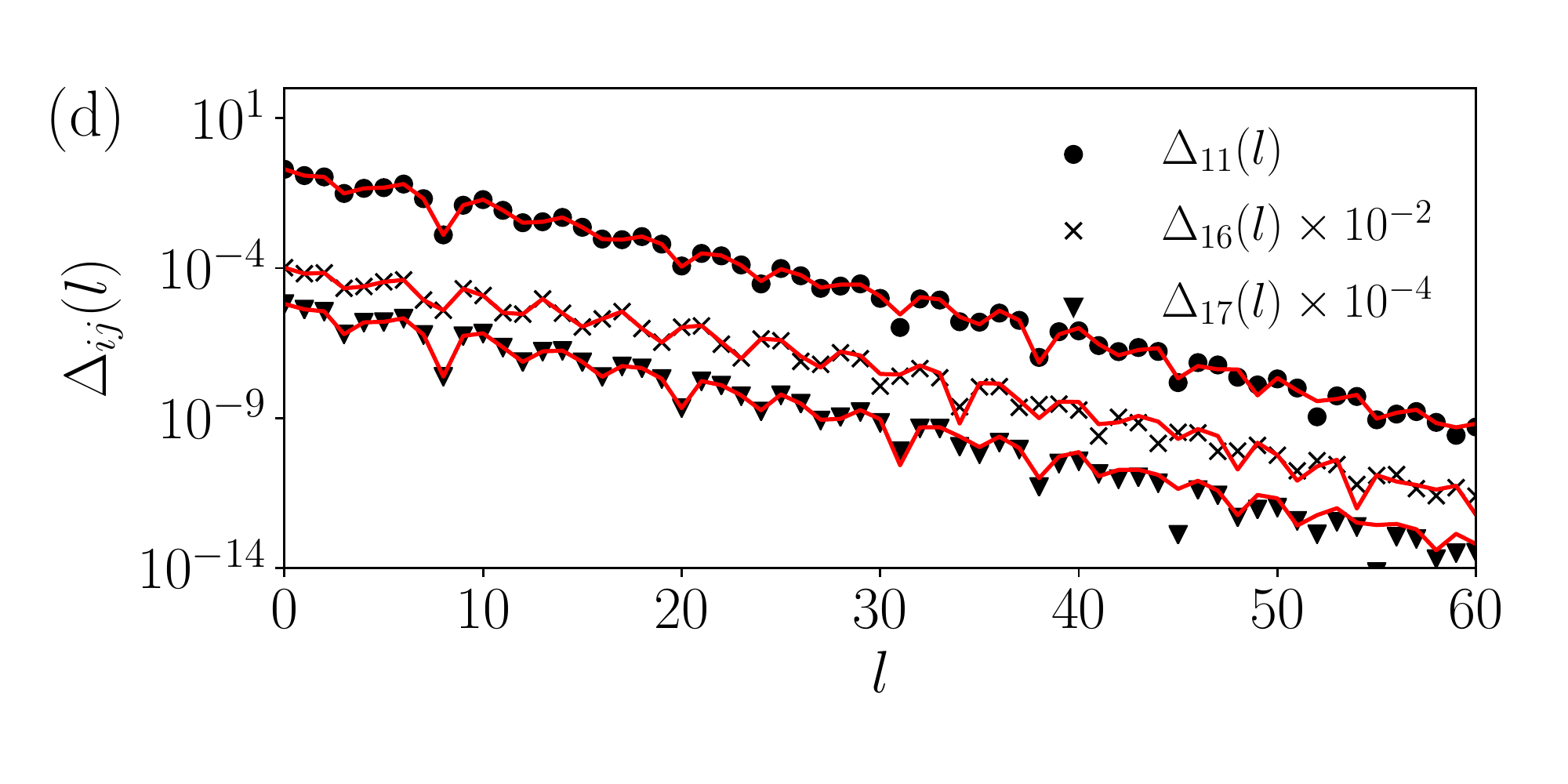}
  \caption{
  (Color online) 
  (a) Crystal structure of LaFeAsO. Four Fe atoms are shown.
  (b) Band structure computed from the tight-binding model. Energy is measured from the Fermi level.
  (c) Convergence of the relative residual norm for the hybridization function for $\beta=500$ eV$^{-1}$ ($T\simeq 24$ K).
  (c) Comparison of the exact (markers) and fitted (lines) hybridization functions for $\Nb=332$. The indices 1, 6, and 7 correspond to the $d_{3Z^2-1}$ orbital on Fe1, the $d_{3Z^2-1}$ orbital on Fe2, and the $d_{XY}$ orbital on Fe2, respectively [see (a)].
  }
  \label{fig:Fe}
\end{figure}
\textit{Results of realistic model for Fe-based superconductors}--
The present algorithm was benchmarked for a realistic five-orbital 2$\times$2 cluster impurity model for LaAsFeO.
The crystal structure is shown in Fig.~\ref{fig:Fe}(a).
A unit cell contains one Fe atom.
Here, the tight-binding model constructed in Ref.~\cite{Kuroki2008}, where each Fe atom has five 3$d$ orbitals, was adopted.
Both the orbital and short-ranged antiferromagnetic correlations play an essential role in Fe-based high-$T_\mathrm{c}$ compounds~\cite{Dai2015}; therefore, a quantitative prediction of the future $T_\mathrm{c}$ may require at least cluster DMFT calculations using a $2\times 2$ supercell in the $ab$ plane.
They require the solution of a 20-orbital impurity model (the number of spin orbitals is 40).

A spin-diagonal matrix-valued hybridization function in the noninteracting limit was constructed using the standard procedure of cluster DMFT at $\beta=500$ eV$^{-1}$ ($T\simeq 24$ K $< T_\mathrm{c}\simeq 26$ K).
Each spin sector of the hybridization function is a $20\times 20$ matrix-valued function at each Matsubara frequency.
It was transformed into the IR basis ($\wmax=20$) and fitted via the procedure used for the random models.

Figure~\ref{fig:Fe}(c) plots $\Nb$ per spin versus the relative residual norm of the fit.
The residual decays quickly with respect to $\Nb$.
The quality of the fit for $\Nb=332$ is assessed by plotting the hybridization function in Fig.~\ref{fig:Fe}(d).
For both inter-atom and intra-atom components,
the discretized model approximately fits the hybridization with four significant digits.
Considering the spin degrees of freedom,
for $\Nb=332$, solve a discretized model with a total of 40+$\Nb$=372 spin orbitals must be solved.
A recently developed exact diagonalization solver with the truncation of the Hilbert space can handle $\Nb > 300$~\cite{Lu:2014de}, but is limited to zero $T$ and a single impurity orbital.
Extensions to finite $T$ and multi impurity orbitals are necessary to solve such a large realistic model.

The group LASSO can be applied to bath fitting in the real-frequency formalism.
Comparisons with nonuniform meshes used in the numerical renormalization group (NRG)~\cite{Bulla:2008bn} and an efficient exact-diagonalization solver~\cite{Lu:2014de} may produce interesting results.
The present algorithm can be combined with a recently proposed bath compression method
which  requires the construction of an accurate discretized model in advance~\cite{Nusspickel2020}.
The present algorithm can be used for zero-$T$ quantum embedding calculations by introducing a fictitious temperature.

In this letter, an efficient and stable discretization algorithm for large-scale impurity models with low symmetries is proposed. The proposed algorithm uses a regularization term based on group LASSO, a sparse-modeling technique, as well as a compact representation of Matsubara Green's function. Its efficiency was demonstrated for random models with several spin orbitals $\Nimp \gg 10$.
We revealed that the required number of bath levels scales only linearly with $\Nimp$.
Additionally, the number of required bath levels for a 20-impurity model for LaAsFeO was estimated.
These results encourage future quantum embedding simulations of real materials,
and set quantitative goals for the future development of classical and quantum algorithms for large-scale impurity problems.
%

HS thanks Markus Wallerberger for the critical reading of the manuscript and
useful comments.
Part of the calculations was run on the facilities of the
Supercomputer Center at the Institute for Solid State Physics,
University of Tokyo.
This research was conducted using the Fujitsu PRIMERGY CX400M1/CX2550M5 (Oakbridge-CX) in the Information Technology Center, The University of Tokyo.
We used the \texttt{irbasis} library~\cite{irbasis2019} for computing IR basis
functions. We used DCore~\cite{DCore} based on TRIQS~\cite{Parcollet:2015gq} and TRIQS/DFTTools~\cite{Aichhorn:2016cd} for computing the hybridization function for LaFeAsO.
HS was supported by JSPS KAKENHI Grant Nos. 18H01158 and 16K17735.
YN was partially supported by JSPS-KAKENHI Grant Numbers 18K11345.
We used VESTA 3~\cite{Momma:2011dd} for visualizing the crystal structure.

\bibliography{ref,aux}

\clearpage
\pagebreak
\begin{center}
\textbf{\large Supplemental Material: Sparse modeling of large-scale quantum impurity models with low symmetries}
\end{center}
\setcounter{equation}{0}
\setcounter{figure}{0}
\setcounter{table}{0}
\setcounter{page}{1}
\makeatletter
\renewcommand{\theequation}{S\arabic{equation}}
\renewcommand{\thefigure}{S\arabic{figure}}
\renewcommand{\bibnumfmt}[1]{[S#1]}
\renewcommand{\citenumfont}[1]{S#1}

\section{Derivative of the cost function}
To constrain $\epsilon_b$ in the interval of $[-\wmax, \wmax]$,
we parameterize $\epsilon_b$ as 
\begin{align}
    \epsilon_b &= \wmax \cos \theta_b.
\end{align}

We give the explicit forms of the derivatives of the cost functions with respect to $\theta_b$ and $V_{bi}$.
To simplify notation, we define
\begin{align}
  g_{lb} &\equiv -S_l^\mathrm{F} V_l(\epsilon_b),\\
  D_{lij} &\equiv \Delta_{ij}(l) - \sum_b V_{bi}^* g_{lb} V_{bj}.
\end{align}
Then, the explicit forms of the derivatives of the cost function is given by
\begin{align}
    \frac{\partial f}{\partial \theta_b} &= \left(\frac{\partial f}{\partial \epsilon_b}\right) \wmax \cos \theta_b,\\
    \frac{\partial f}{\partial \mathrm{Re}~V_{bi}} &= 2 \mathrm{Re}\left(\frac{\partial f}{\partial V_{bi}} \right),\label{eq:df-Re}\\
    \frac{\partial f}{\partial \mathrm{Im}~V_{bi}} &= -2 \mathrm{Im}\left(\frac{\partial f}{\partial V_{bi}} \right),\label{eq:df-Im}
\end{align}    
where
\begin{align}
    \frac{\partial f}{\partial \epsilon_{b}} &= -2 \mathrm{Re} \sum_{lij} V_{bi} \left(\frac{\partial g_{lb}}{\partial \epsilon_b}\right)V_{bj}^*,\\
  \frac{\partial g_{lb}}{\partial \epsilon_a} &=
  -\sqrt{\beta/2}~s_l^\mathrm{F} \wmax^{-1} (v_l^\mathrm{F})^\prime (\epsilon_b/\wmax),\\
  \frac{\partial f}{\partial V_{bk}} &=
  -\sum_{jl}g_{lb}^* V_{bj}^* D_{lkj} - \sum_{il} g_{lb} V_{bi}^* D_{lik}^*.\label{eq:df-Vbk}
\end{align}
In Eqs.~(\ref{eq:df-Re}), (\ref{eq:df-Im}), (\ref{eq:df-Vbk}),
$\frac{\partial f}{\partial V_{bi}}$ must be regarded as Wirtinger derivative.

\section{Optimization algorithm}
\begin{figure}[h!]
\begin{algorithm}[H]
        \begin{algorithmic}
                \State $\boldsymbol{\epsilon} \gets \text{Non-uniform~grid points~from~IR~basis}$
                \State $\theta_b \gets \cos^{-1}(y_b/\wmax)$
                \State $\boldsymbol{V} \gets \text{Random~numbers~from~normalized~Gaussian~distribution}$
                \State Optimize $\boldsymbol{V}$ using L-BFGS method
                \State Optimize $\boldsymbol{V}$ and $\boldsymbol{\theta}$ using L-BFGS method
                \State Compute $\|v_b\|$ and sort the bath levels in descending order
                \State $r_0 \gets \text{relative residual norm}$
                \For{$k = 1, 2, ...$}
                \State $r \gets \text{relative residual norm without the last bath level}$
                \If{$r>1.1\times r_0$}
                \State Exit loop
                \EndIf
                \State Remove the last bath level
                \EndFor
        \end{algorithmic}
        \caption{Optimization algorithm}
        \label{algorithm:opt}
\end{algorithm}
\end{figure}
Algorithm~\ref{algorithm:opt} shows a pseudocode for the optimization algorithm.
In the pseudocode, we use 
$\boldsymbol{\theta}=(\theta_1, \cdots, \theta_{N_\mathrm{b}^0})$,
and $\boldsymbol{\epsilon} = (\epsilon_1, \cdots, \epsilon_{N_\mathrm{b}^0})$ and $\boldsymbol{V}=(V_{11}, \cdots, V_{1\Nimp},\cdots)$.

\section{Robustness against noise}
To test the robustness of the present algorithm against in the hybridization function,
we add Gaussian noise to $\Delta_{ij}(l)$ used to procedure the data in Fig. 2 of the main text.
The standard deviation of the noise is chosen to be $10^{-5}$.
We fit the hybridization function with the noise using exactly the same procedure.
A remarkable different from the result without noise is the existence of a \textit{overfitting} regime ($\alpha < 10^{-3}$),
which is signaled a flat region of $\delta(\alpha)$ and the jump in $\Nb(\alpha)$.
The algorithm works efficiently down to $\alpha=10^{-3}$.

\begin{figure}[h!]
  \centering
  \includegraphics[width=0.9\linewidth]{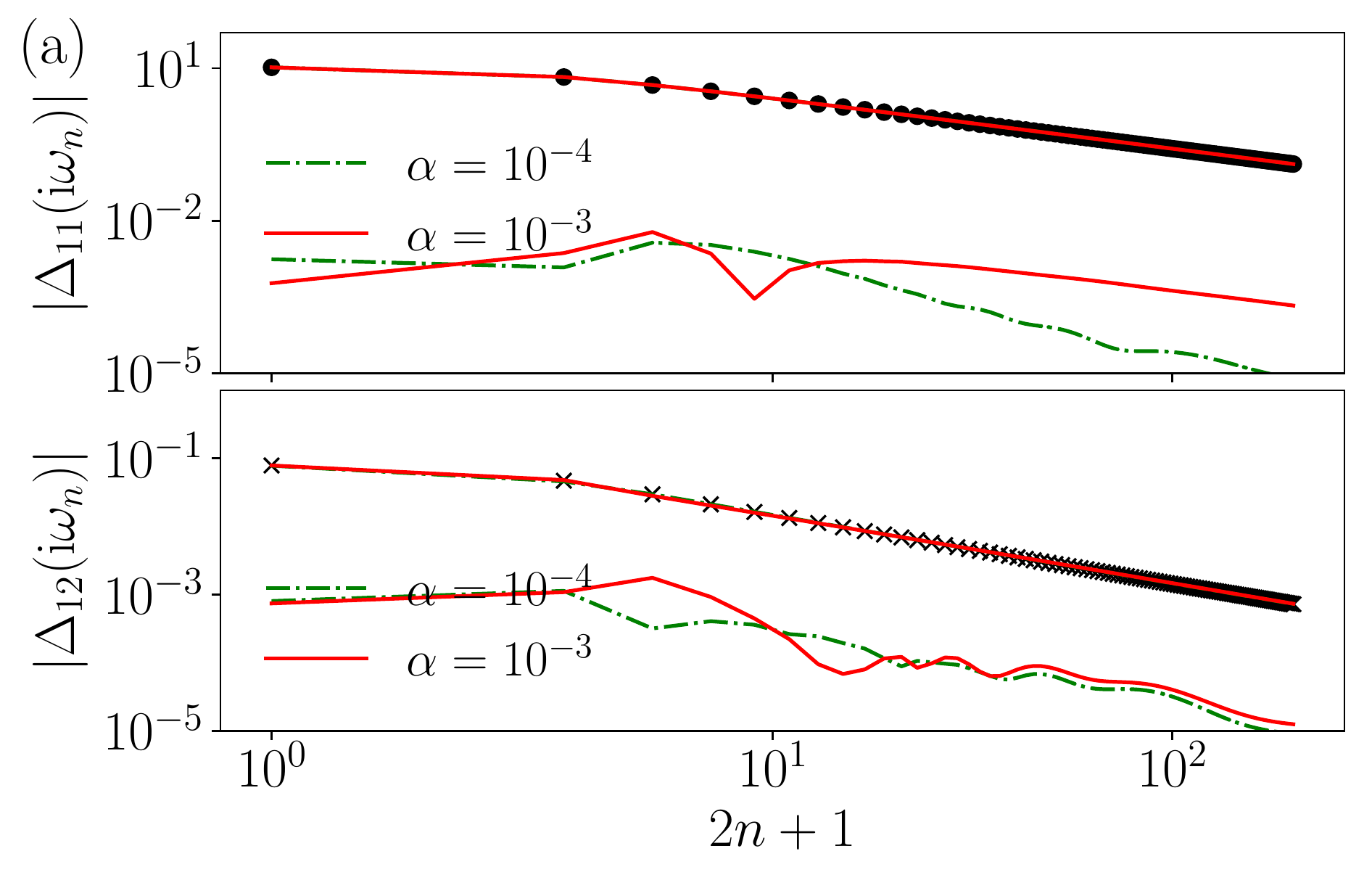}
  \includegraphics[width=0.9\linewidth]{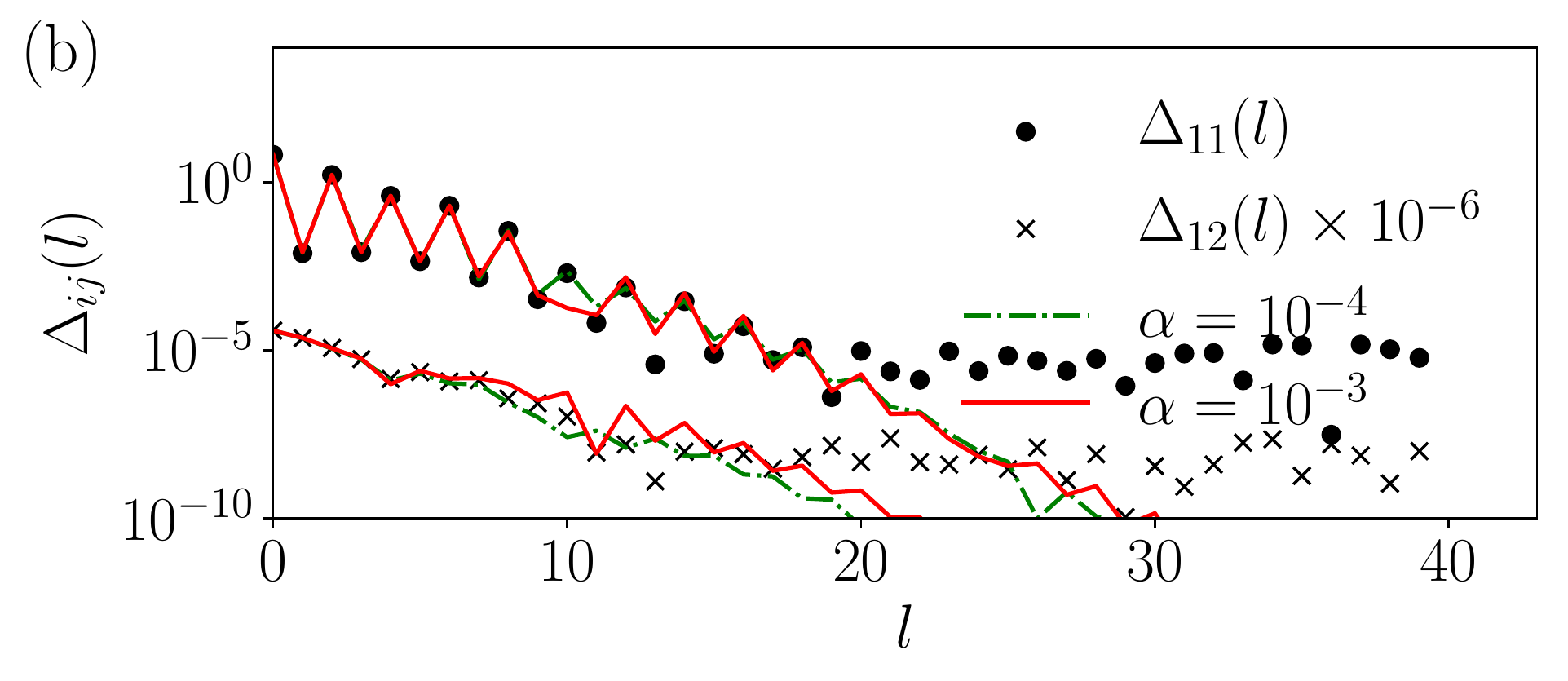}
  \includegraphics[width=0.9\linewidth]{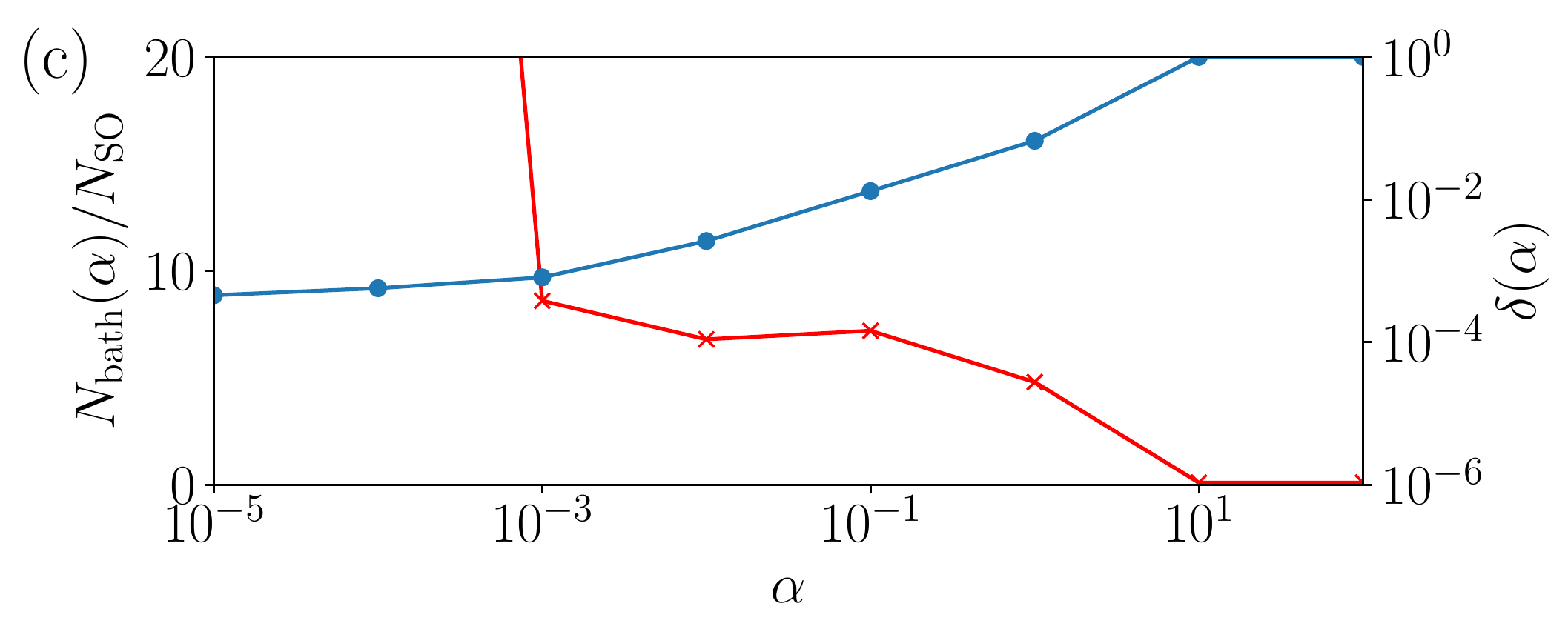}
  \includegraphics[width=0.9\linewidth]{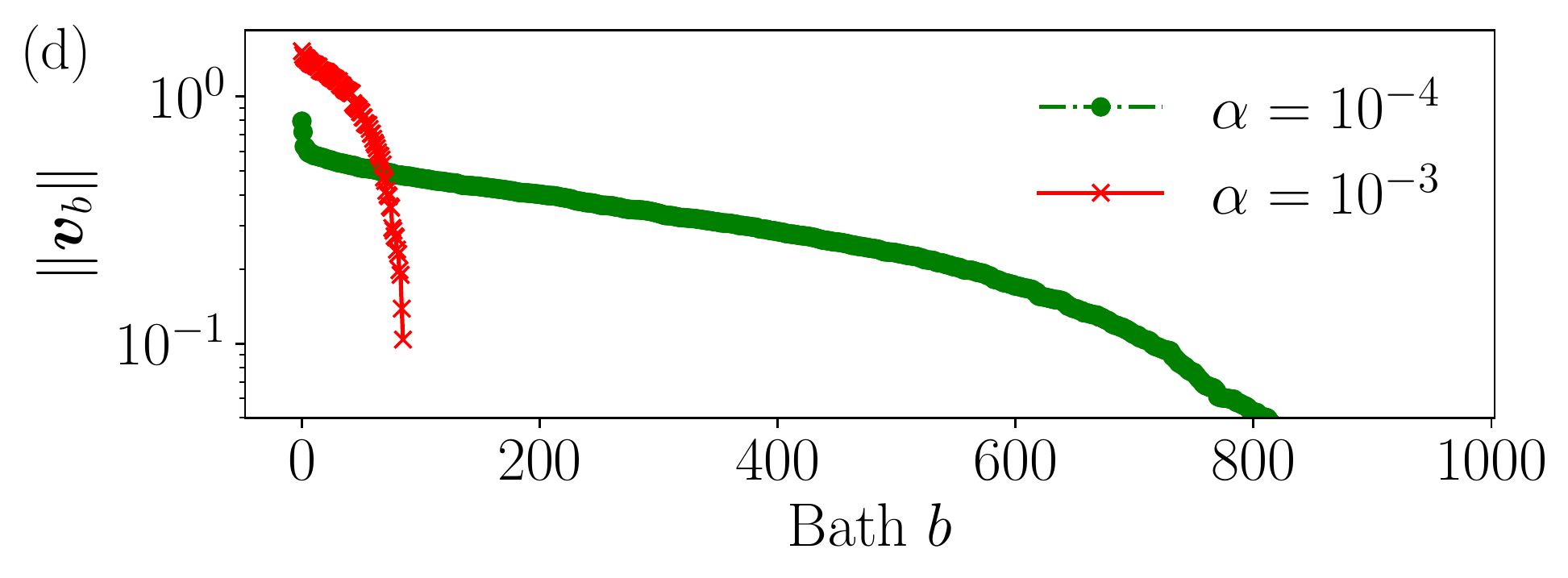}
  \includegraphics[width=0.9\linewidth]{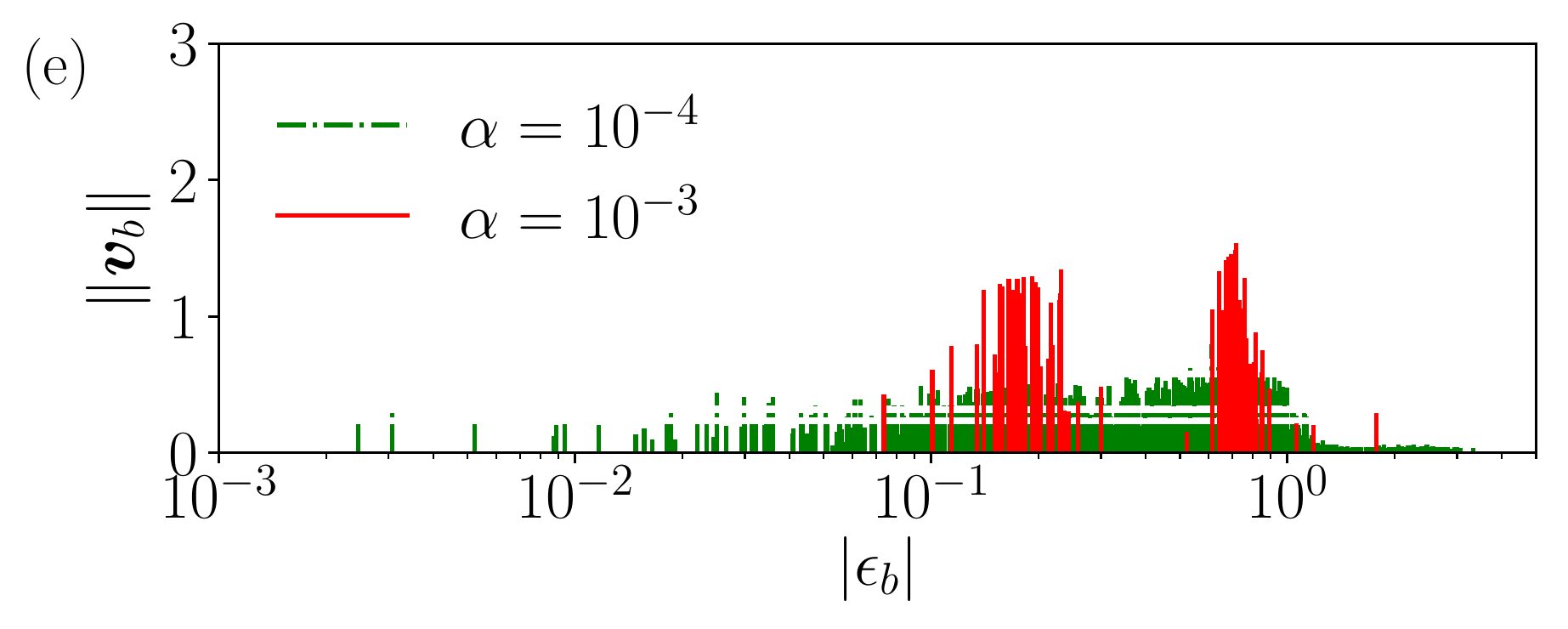}
  \vspace{-1em}
  \caption{
  (Color online) 
  Results of one sample for $\Nimp=10$ and $\beta=10$ and the noise level of $10^{-5}$.
  The hybridization function and the fitted results are compared in Matsubara frequency [(a)] and in the IR basis [(b)], respectively.
  In (a) and (b), the fitting error is also shown.
  (c) Number of bath levels $\Nb$ and relative residual norm $\delta$.
  (d) Strength of the coupling to the impurity $\|\vecv_b\|$ in the decreasing order. (e) Positions of $\epsilon_b$ and $\|\vecv_b\|$.
  }
  \label{fig:random-noise}
\end{figure}
\end{document}